\documentclass{nature}
\usepackage{xcolor, ulem, lineno, pdfpages, caption}
\usepackage{amssymb, amsmath, txfonts, textcomp, marvosym, upgreek}
\usepackage{aas_macros, graphicx, pdflscape, hyperref, verbatim}
\usepackage{longtable, threeparttable, threeparttablex, adjustbox, booktabs, url, subfigure}
\usepackage{multibib}%\usepackage[resetlabels]{multibib}
\usepackage{caption}
\usepackage{footnote} % 加载 footnote 宏包
\makesavenoteenv{table} % 允许在表格环境中使用脚注
\newcites{md}{Reference}
\makeatletter
\usepackage[symbol]{footmisc}

\def\thanks#1{\protected@xdef\@thanks{\@thanks\protect\footnotetext{#1}}}
\let\saved@includegraphics\includegraphics
\AtBeginDocument{\let\includegraphics\saved@includegraphics}

\makeatother
\let\oldequation\equation
\let\oldendequation\endequation
\renewenvironment{equation}{\linenomathNonumbers\oldequation}{\oldendequation\endlinenomath}
\usepackage{multirow}

\title{Pulsed radio emission from a Central Compact Object}

\author{
Lei Zhang$^{1,2}$\href{https://orcid.org/0000-0001-8539-4237}{\includegraphics[scale=0.07]{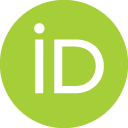}}
Alessandro Ridolfi$^{3}$\href{https://orcid.org/0000-0001-6762-2638}{\includegraphics[scale=0.07]{ORCIDiD.png}},
Di Li$^{4}$\href{https://orcid.org/0000-0003-3010-7661}{\includegraphics[scale=0.07]{ORCIDiD.png}}\textsuperscript{\Letter}\thanks{Email: dili@tsinghua.edu.cn (D.L.)},
Erbil G\"{u}gercino\u{g}lu$^{5}$,
Fernando Camilo$^{6}$\href{https://orcid.org/0000-0002-1873-3718}{\includegraphics[scale=0.07]{ORCIDiD.png}},\\
Wynn C. G. Ho$^{7}$\href{https://orcid.org/0000-0002-6089-6836}{\includegraphics[scale=0.07]{ORCIDiD.png}},
Matthew Bailes$^{2}$\href{https://orcid.org/0000-0003-3294-3081}{\includegraphics[scale=0.07]{ORCIDiD.png}},
Ping Zhou$^{8,9}$\href{https://orcid.org/0000-0002-5683-822X}{\includegraphics[scale=0.07]{ORCIDiD.png}},
Craig O. Heinke$^{10}$\href{https://orcid.org/0000-0003-3944-6109}{\includegraphics[scale=0.07]{ORCIDiD.png}},
Marcus E. Lower$^{2}$\href{https://orcid.org/0000-0001-9208-0009}{\includegraphics[scale=0.07]{ORCIDiD.png}}
}

\begin{document}
%\linenumbers
\maketitle
%\footnotemark[1]
%\footnotetext[1]{These authors contributed equally to this work.}
\begin{affiliations}
\item {State Key Laboratory of Radio Astronomy and Technology, National Astronomical Observatories, Chinese Academy of Sciences, Beijing 100101, China}
\item {Centre for Astrophysics and Supercomputing, Swinburne University of Technology, Hawthorn, VIC 3122, Australia}
\item {Fakult\"at f\"ur Physik, Universit\"at Bielefeld, Postfach 100131, Bielefeld {\rm D-33501}, Germany}
\item {New Cornerstone Science Laboratory, Department of Astronomy, Tsinghua University, Beijing 100084, China}
\item {School of Arts and Sciences, Qingdao Binhai University, Huangdao District, Qingdao {\rm 266555}, China}
\item {South African Radio Astronomy Observatory, Liesbeek House, River Park, Gloucester Road, Cape Town, 7700, South Africa}
\item {Department of Physics and Astronomy, Haverford College, 370 Lancaster Avenue, Haverford, PA 19041, USA}
\item {School of Astronomy and Space Science, Nanjing University, Nanjing 210023, China}
\item {Key Laboratory of Modern Astronomy and Astrophysics, Nanjing University, Ministry of Education, Nanjing 210023, China}
\item {Department of Physics, CCIS 4-183, University of Alberta, Edmonton, AB T6G 2E1, Canada}
\end{affiliations}
\vspace{0.2in}

\begin{abstract}
Located at the centres of supernova remnants, central compact objects (CCOs) are among the most puzzling neutron stars. CCOs are bright in thermal X-rays, yet have evaded detection by major radio telescopes for decades, giving rise to the view that they are intrinsically radio-quiet and possess exceptionally weak magnetic fields. Here we show that the prototypical young CCO 1E 1207.4–5209 is in fact a faint radio pulsar rotating at the 0.4s X-ray period. Analysis of its polarization indicates that the radio beam intersects our line of sight near the magnetic pole, affirming its radio faintness’ being intrinsic. Once its supernova remnant dissipates, this source would be misidentified as an apparently gigayear-old pulsar. The CCO's low radio flux density may explain why many supernova remnants lack detectable radio pulsars and suggests a hidden population of young, slowly rotating neutron stars.
\end{abstract}
%The summary paragraph more than 200 words.
%https://www.nature.com/documents/nature-summary-paragraph.pdf
%The typical length of a 6-page article with 4 modest display items (figures and tables) is 2500 words (summary paragraph plus body text). 
% Let's keep the US spelling

%\section*{Central compact objects and the origin of their radio silence}
Young neutron stars manifest themselves in diverse forms, including rotation-powered pulsars, magnetars, and the enigmatic class of central compact objects (CCOs). CCOs, with ten confirmed members\footnote{\url{http://www.iasf-milano.inaf.it/~deluca/cco/main.htm}\label{web:CCOs}}, have  been defined as point-like X-ray sources at the geometrical centres of supernova remnants (SNRs), without optical counterparts, radio emission, or associated pulsar wind nebulae (PWNe)~\cite{DeLuca2017}. 

Over the past two decades, searches for radio pulses from CCOs have yielded only upper limits~\cite{Gaensler2000, Halpern2007, Lu2024}. This persistent radio silence, along with purely thermal X-ray emission and their exceptionally low spin-down rates, have led to the hypothesis that CCOs may represent a distinct evolutionary pathway among young neutron stars~\cite{Wu2021}. The low B fields of CCOs inspired the ``anti-magnetar” model~\cite{Halpern2010, Gotthelf2013}, in which CCOs are born with intrinsically weak external dipole fields. One explanation for this involves magnetic-field burial by fallback accretion~\cite{Ho2011, Vigano2012}, where initially strong fields are submerged beneath the crust by a large amount of mass accreted soon after neutron star formation ~\cite{Chevalier1989,Muslimov1996,Geppert1999,Bernal2010} and gradually re-emerge over timescales of $10^3$–$10^5$ years. Another possibility invokes a stochastic dynamo during the proto-neutron star phase~\cite{Gourgouliatos2020, Igoshev2021}, which generates a highly tangled internal field, with either a weak or absent dipole component. In addition to neutron star evolution, geometrical factors, such as unfavorable beam orientation or a narrow inclination angle between the rotation and magnetic dipole axes, can prevent radio detection.

\section*{Detection of radio pulsations from 1E~1207.4$-$5209}
1E~1207.4$-$5209 is an archetypal CCO. Its associated SNR PKS~1209$-$51/52 (G296.5+10.0) has an estimated age of $\sim$14 kyr~\cite{Eppens2024} and a distance of just $\sim$1.4 kpc~\cite{Eppens2024}. It exhibits 424-ms X-ray pulses~\cite{Zavlin2000}, confirming its neutron-star nature, and unusual absorption features in its spectrum~\cite{Sanwal2002, Bignami2003}, interpreted as cyclotron or atomic lines. The agreement of the dipolar B field ($\sim8\times10^{10}$ G) inferred from the cyclotron line interpretation \cite{Bignami2003}, and the dipolar B field ($\sim9.8\times10^{10}$ G) inferred from its period and period derivative \cite{Gotthelf2007, Gotthelf2013}, established 1E~1207.4$-$5209 as a low magnetic field neutron star. 

Using the MeerKAT array in the UHF band (544–1088\,MHz), we conducted a targeted search for radio pulsations from 1E~1207.4$-$5209 and three others (See Extended Data Table~\ref{tab:4CCOobs}). We detected coherent radio emission with a period of 424\,ms and a dispersion measure (DM) of 69\,pc cm$^{-3}$ at a signal-to-noise ratio of $\approx 20$ in the discovery observation, independently confirmed by both time-domain and Fourier-domain periodicity search pipelines (Extended Data Figure~\ref{fig:FFA-FFT})—a period that matches that of the X-ray pulsar PSR~J1210$-$5226 and a DM consistent with the estimated distance to its host supernova remnant, PKS~1209$-$51/52 (DM models give distance estimates of 1.9 or 1.5 kpc, compared with 1.4 kpc from X-ray studies of the SNR; see Table~\ref{tb:1E1207_par}). This confirms that at least some CCOs can operate as active radio pulsars. We measured a rotation measure (RM) of $10 \pm 1$\,rad\,m$^{-2}$ and estimate a mean flux density of approximately 21\,$\mu$Jy at 816\,MHz. Archival MeerKAT 1.3\,GHz radio continuum images of PKS~1209$-$51/52 show no evidence of bright radio emission at the pulsar’s location~\cite{Cotton2024}, and the TRAPUM pulsar search targeting this remnant reported no detections, placing a limiting flux density of $\sim$21\,$\mu$Jy at 1284\,MHz~\cite{Turner2024}.

To further characterize its emission, we obtained an 8.5-hour dual-frequency MeerKAT observation in split-array mode at UHF and L-band. The pulse flux varied significantly over time and frequency (Methods and Extended Data Fig.~\ref{fig:DDT-FFT}), consistent with interstellar scintillation typical of pulsars at that DM ~\cite{Gitika2023}, which may have contributed to previous non-detections. Figure~\ref{fig:1E1207_PolProf} shows the integrated polarization profiles from both bands, and Table~\ref{tb:1E1207_par} summarizes key radio and X-ray parameters.

\section*{Radio properties}
The radio properties of PSR~J1210$-$5226 provide a unique opportunity to assess how CCOs relate to the broader pulsar population. The measured profile width $W_{10}$ of $\sim$17\% is relatively wide but within the observed range for ordinary pulsars with similar spin periods~\cite{Posselt2021}. Its linear polarization fraction however ($\sim$77\%) is unusually high for pulsars with such low spin-down energy, and is more typical of young, high-$\dot{E}$ systems~\cite{Wang2023}, perhaps suggesting a dynamically young or recently reconfigured magnetosphere. Given the pulsar's characteristic age of $\sim$300 Myr~\cite{Perez2025} compared to the $\sim$14 kyr age of the associated supernova remnant~\cite{Eppens2024}, the current spin period is likely close to its birth value.

Further constraints on the pulsar geometry are provided by a Rotating Vector Model (RVM) fit to the polarization position angle swing. The data yield relatively weak constraints due to the limited signal-to-noise ratio and position angle phase coverage (Figure~\ref{fig:RVMfit} and Methods). Population modeling suggests pulsar emission originates from $\lesssim 1000$\,km above the surface \cite{Johnston2023}, which restricts the geometric space sampled by the RVM in Figure~\ref{fig:RVMfit} to a magnetic inclination angle ($\alpha$) of $\lesssim 53^{\circ}$ and an impact parameter ($\beta$) of $\lesssim -17^{\circ}$. Such a geometry is consistent with a largely aligned rotator and explains the simple, nearly sinusoidal X-ray pulse profile of 1E~1207.4$-$5209, which is thought to arise from thermal emission from a single hot spot always in view~\cite{Zavlin2000, Bignami2003, Gotthelf2013}.

The radio emission itself appears consistent with ordinary pulsar-like behaviour despite the unusually weak inferred dipole field. The pulse morphology and polarization do not exhibit the strong variability or complex swings observed in more exotic neutron-star systems. For example, magnetar radio emission is often transient and associated with magnetospheric reconfiguration following X-ray outbursts, while polarization variations in some repeating fast radio bursts have been attributed to rapidly evolving magnetospheric or propagation effects~\cite{McKinven2025, Liu2025}. In contrast, the emission from 1E~1207.4$-$5209 appears relatively stable.

PKS~1209$-$51/52~\cite{Cotton2024} is a bilaterally symmetric supernova remnant \textcolor{blue}{(Extended Data Fig.~\ref{fig:MKT-Xray})}. Radio polarization observations reveal a distinct east-west RM asymmetry in the remnant, with mean values of +28 rad m$^{-2}$on the eastern limb and $-14$ rad m$^{-2}$ on the western limb. The foreground RM along the line of sight to the SNR is estimated to lie in the range of +10 to +20 rad m$^{-2}$ \cite{harvey-smith10}. The RM value associated with the CCO ($+$10 rad m$^{-2}$) is consistent with the range toward the supernova remnant, the range toward the supernova remnant, albeit with considerable uncertainty for the latter measurements. This consistency provides no evidence of a strong magneto-ionic environment surrounding the CCO, such as one that could be produced by a magnetized fallback disk. This agrees with HST and VLT measurements, which suggest an ultra-low mass ($<10^{-10}~M_\odot$) for any potential fallback disk surrounding this CCO \cite{deluca2011}.

In addition to 1E~1207.4$-$5209, we observed three other CCOs (RX~J0822.0$-$4300, CXOU J085201.4$-$461753, and 1WGA J1713.4$-$3949) in our MeerKAT campaign but did not detect pulsed radio emission, placing flux-density limits of approximately 9~$\mu$Jy at 816 MHz (Methods). These non-detections may reflect beaming effects~\cite{Tauris1998, Kolonko2004}, interstellar scintillation, intrinsically low luminosities, or limited sensitivity~\cite{Zhang2019b, Zhang2023}. It is also possible that 1E~1207.4$-$5209 may represent a rare transitional phase in which re-emerging magnetic fields briefly enable coherent radio emission. 

The detection follows a previously reported glitch episode~\cite{Gotthelf2018, Gotthelf2020}, suggesting that changes in the star’s rotational or magnetic configuration may influence the activation or detectability of radio emission. The absence of narrow features in the X-ray pulse profile may further support a low magnetic-field interpretation, although the field configuration remains uncertain. Overall, the emission profile, polarization properties, and inferred geometry indicate that PSR~J1210$-$5226, while radio faint, exhibits characteristics broadly consistent with ordinary pulsars~\cite{Everett2001}.

\section*{Implications for CCO magnetic field evolution}
Previous searches for radio pulsations from 1E~1207.4$-$5209 and other CCOs using sensitive instruments such as the GBT, FAST, and earlier MeerKAT campaigns had returned null results~\cite{Halpern2007, Lu2024, Turner2024}, reinforcing the perception that CCOs are intrinsically radio-quiet. Our detection shows that even a relatively weak dipole field can sustain coherent radio emission, challenging the long-standing view that they are intrinsically radio inactive.

Two broad explanations may account for this detection. One involves magnetic-field topology: a complex internal field configuration with highly multipolar structure, rather than a uniformly weak dipole field, could reconcile the properties of 1E~1207.4$-$5209 with the persistent radio silence of other CCOs. As discussed in the Methods, a stochastic dynamo during the proto-neutron star phase may generate a tangled magnetic field residing within the crust, the evolution of which allows for localized multipolar magnetic structures, providing conditions favorable for radio emission. The second possibility invokes magnetic-field evolution provoked by fallback accretion: 1E~1207.4$-$5209, with its longer spin period and possibly older age compared to the other two X-ray pulsating CCOs, may have allowed sufficient time for a previously buried dipole field~\cite{Chevalier1989,Bernal2010} to re-emerge~\cite{Muslimov1996,Geppert1999,Ho2011,Ho2015}, potentially accelerated by the 2015 glitch~\cite{Gotthelf2018,Gotthelf2020,Perez2025}. The timescales for magnetic restructuring are long compared to existing observations. 
To distinguish these scenarios, continued radio monitoring, particularly measurements of spin-down evolution, will be essential.

\section*{Implications for the population of radio neutron stars}
Once the supernova remnant has dissipated, PSR J1210--5226 may well continue to be an active radio pulsar for a Gyr or more and be indistinguishable from the normal pulsar population, albeit of low luminosity. It gives support to suggestions\cite{Narayan1987,Gullon2014} that some or indeed a high fraction of neutron star pulsars are ``injected'' into the population with birth spin periods comparable to the median period of the observed population, ie $\sim$0.5 s. The low radio luminosity of PSR~J1210--5226 further suggests that a significant population of similar objects will remain undetected, potentially explaining the absence of detectable radio emission from neutron stars in other SNRs and binary systems.

Our detection of PSR~J1210$-$5226 establishes the first confirmed radio-emitting CCO, providing a direct observational link between this class of thermal X-ray sources and the broader pulsar population. 
It demonstrates that CCOs are not an entirely radio-silent population, but may instead represent a class of intrinsically faint or intermittently detectable radio pulsars. Continued monitoring of PSR~J1210$-$5226, together with deeper and more sensitive searches for radio emission from other CCOs, will be crucial to determine whether radio activity is a common but elusive property of the class or a rare characteristic of this source. These results underscore the importance of high-sensitivity surveys with instruments such as MeerKAT and future facilities like the Square Kilometre Array in uncovering the full diversity of the Galactic neutron star population.

\begin{table*}
\centering
\caption{Parameters for PSR~J1210$-$5226}
\label{tb:1E1207_par}
\renewcommand{\arraystretch}{0.56}
\setlength{\tabcolsep}{0.3mm}{
\begin{tabular}{lllll}
\hline
\multicolumn{3}{l}{Parameter}                               & \multicolumn{2}{c}{Value} \\ \hline
\multicolumn{5}{c}{Parameters from MeerKAT (544--1088\,MHz)}\\ \hline        
\multicolumn{3}{l}{Right Ascension (J2000)$^{a}$}           & \multicolumn{2}{c}{$12^{\rm h}\;10^{\rm m}\;00^{\rm s}.91$} \\   
\multicolumn{3}{l}{Declination (J2000)$^{a}$}               & \multicolumn{2}{c}{$-52^{\circ}\;26^{'}\;28^{''}.4$}\\
\multicolumn{3}{l}{Period, $P$ (s)}                         & \multicolumn{2}{c}{0.4241307(1)} \\
\multicolumn{3}{l}{Predicted $P$ from X-ray timing$^{b}$ (s)} & \multicolumn{2}{c}{0.4241307605(1)} \\
\multicolumn{3}{l}{Dispersion measure, DM (pc cm$^{-3}$)}   & \multicolumn{2}{c}{69.0(4)} \\
\multicolumn{3}{l}{Rotation Measure, RM (rad m$^{-2}$)}     & \multicolumn{2}{c}{$+$10(1)} \\
\multicolumn{3}{l}{Epoch of ephemeris (MJD)}                & \multicolumn{2}{c}{60314.961} \\
\multicolumn{3}{l}{Pulse width at 50\% of peak, $W_{50}$ ($^{\circ}$)}   & \multicolumn{2}{c}{33} \\
\multicolumn{3}{l}{Pulse width at 10\% of peak, $W_{10}$ ($^{\circ}$)}   & \multicolumn{2}{c}{61} \\
\multicolumn{3}{l}{Mean flux density at 816 MHz ($\mu$Jy)}        & \multicolumn{2}{c}{33} \\
\multicolumn{3}{l}{Percentage linear polarization ($L/I$)}   & \multicolumn{2}{c}{77(5)} \\
\multicolumn{3}{l}{Percentage circular polarization ($V/I$)} & \multicolumn{2}{c}{$+$30(4)} \\
\multicolumn{3}{l}{Percentage absolute circular polarization ($\left|V\right|/I$)} & \multicolumn{2}{c}{32(4)} \\
\hline  

\multicolumn{5}{c}{Parameters from X-ray observations\cite{Perez2025}}\\ \hline
\multicolumn{3}{l}{Right Ascension (J2000)}     & \multicolumn{2}{c}{$12^{\rm h}\;10^{\rm m}\;00^{\rm s}.9126(29)$} \\   
\multicolumn{3}{l}{Declination (J2000)}         & \multicolumn{2}{c}{$-52^{\circ}\;26^{'}\;28^{''}.303(42)$}\\
\multicolumn{3}{l}{Period, $P$ (s)$^{c}$}       & \multicolumn{2}{c}{0.424130756710(26)}\\ 
\multicolumn{3}{l}{Period derivative, $\dot{P}$ (s s$^{-1}$)$^{c}$}      & \multicolumn{2}{c}{$2.015(49)\times10^{-17}$} \\ 
\multicolumn{3}{l}{Epoch of ephemeris (MJD)}                             & \multicolumn{2}{c}{58144.000}\\
\multicolumn{3}{l}{Surface dipole magnetic field, $B_{s}$ (G)$^{d}$}     & \multicolumn{2}{c}{$9.8\times10^{10}$}\\ 
\multicolumn{3}{l}{Spin-down luminosity, $\dot{E}$ (erg s$^{-1}$)$^{d}$} & \multicolumn{2}{c}{$1.1\times10^{31}$}\\ 
\multicolumn{3}{l}{Characteristic age, $\tau_{c}$ (Myr)$^{d}$}           & \multicolumn{2}{c}{303}\\ 
\hline

\multicolumn{5}{c}{Distance Estimates}\\ \hline
\multicolumn{3}{l}{From SNR PKS 1209$-$51/52 (kpc)\cite{Eppens2024}} & \multicolumn{2}{c}{1.4} \\
\multicolumn{3}{l}{From DM: NE2001 / YMW16 (kpc)\cite{Cordes2002, Yao2017}} & \multicolumn{2}{c}{1.9 / 1.5} \\
\hline

\multicolumn{5}{l}{{\bf Note.} The radio period is consistent, within uncertainties, with the period extrapolated from the X-ray timing}\\
\multicolumn{5}{l}{ephemeris\cite{Perez2025}, confirming the identification of the radio and X-ray sources. The Stokes parameters were defined}\\
\multicolumn{5}{l}{using the PSR/IEEE convention described in~\cite{vanS10}.}\\
\multicolumn{5}{l}{$^{a}$ The telescope was pointed to the X-ray coordinates, whose position is known to better than $1^{''}$, well within}\\
\multicolumn{5}{l}{~~~the MeerKAT UHF tied-array beam (FWHM $\approx 0.5^{'}$).}\\
\multicolumn{5}{l}{$^{b}$ Predicted using X-ray ephemeris parameters ($P$, $\dot{P}$) at MJD 58144~\cite{Perez2025}.}\\
\multicolumn{5}{l}{$^{c}$ From the post-glitch timing solution (2016-2023).}\\
\multicolumn{5}{l}{$^{d}$ Derived parameters ($B_{s}$, $\dot{E}$, $\tau_{c}$) are based on the pre-glitch timing solution (2002-2014). The characteristic}\\
\multicolumn{5}{l}{~~~age is not indicative of the true age, which is estimated to be $\sim$14 kyr  of its host SNR, PKS~1209$-$51/52\cite{Eppens2024}.}\\
\end{tabular}}
\end{table*}
%\multicolumn{5}{l}{$^{a}$ The positional uncertainty is estimated to be $\sim 0.5'$ along each semi-axis, corresponding to the minimum}\\
%\multicolumn{5}{l}{~~~FWHM of the single tied-array beam in the MeerKAT UHF band.}\\
%0.424130756710+(60314-58144)*86400*(2.015e-17), consider Propagation of Uncertainty 

\begin{figure*}
\centering
\includegraphics[width=0.7\linewidth]{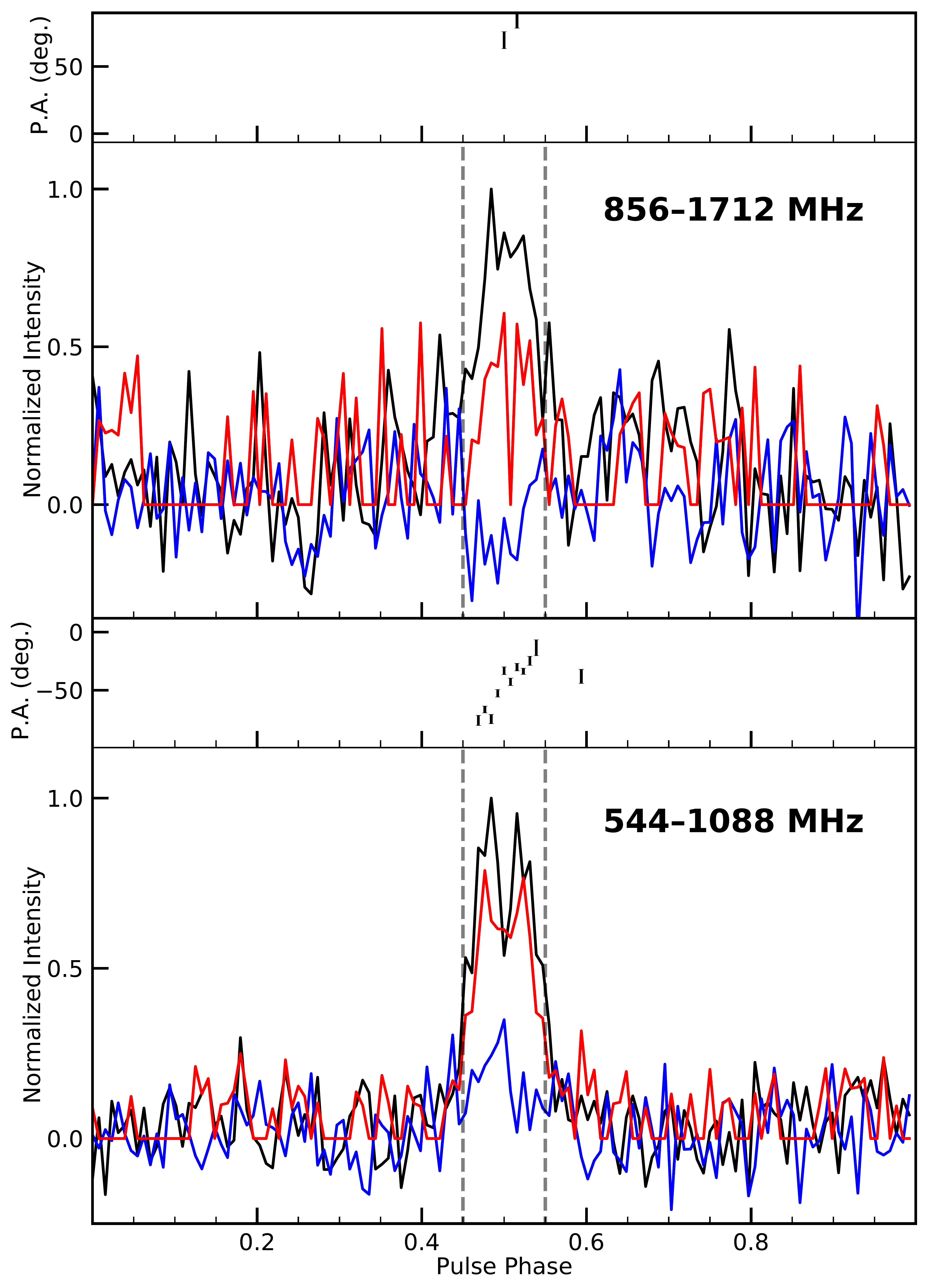}
\caption{{\bf Polarization profiles of PSR~J1210$-$5226.} Integrated polarization profiles from two MeerKAT observations: a 4-hour UHF-band observation (544–1088\,MHz) on 2024 January 5 (bottom), and the first 4.25-hour segment of the L-band observation (856–1712\,MHz) on 2025 October 18 (top). In each panel, black, red, and blue lines represent the total intensity, linear polarization, and circular polarization, respectively. Black points in the upper panels indicate the polarization position angle (PA) of the linear component.}
\label{fig:1E1207_PolProf}
\end{figure*}

\begin{figure*}
\begin{center}
\includegraphics[width=0.8\linewidth]{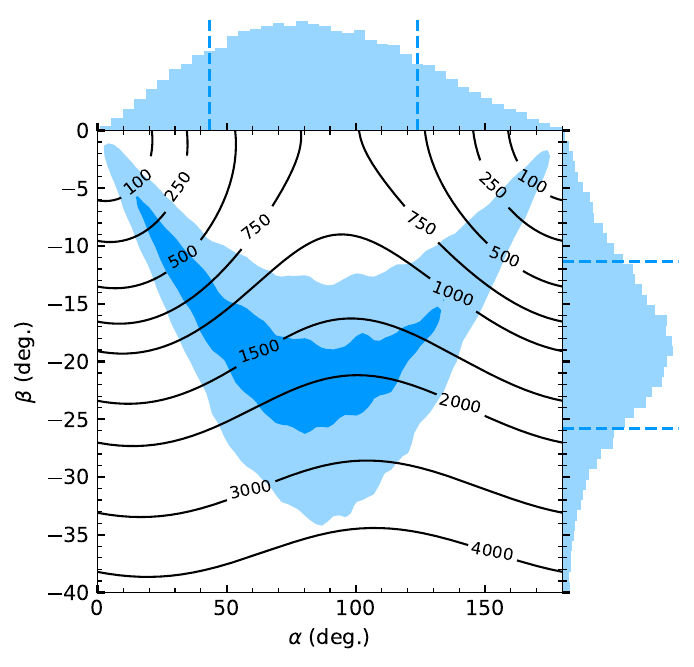}
\end{center}
\caption{{\bf RVM constraints on pulsar geometry.} One- and Two-dimensional posterior distributions of the magnetic inclination angle ($\alpha$) and impact parameter ($\beta$). The central panel shows the 1-$\sigma$ (dark blue) and 2$\sigma$ (light blue) credible intervals. Black contours indicate constant emission heights (km), assuming a filled beam. The marginalized 1-D posterior distributions for $\alpha$ and $\beta$ are shown along the top and right panels respectively, where the dashed lines indicate the 1-$\sigma$ confidence intervals.}
\label{fig:RVMfit}
\end{figure*}

\clearpage
\bibliographystyle{naturemag}
%\bibliography{reference}

\begin{addendum}
\item [Acknowledgments] 
We thank Emma Carli for helpful discussions and Haichen Lin for producing the eROSITA X-ray image.
This work was supported by the National Natural Science Foundation of China (NSFC; grant Nos. 12588202, and 12103069), CAS project No. JZHKYPT-2021-06, and the National Key R\&D Program of China (2023YFE0110500).
This work utilized data from MeerKAT, with data processing conducted primarily on the OzSTAR national facility at Swinburne University of Technology. The MeerKAT telescope is operated by the South African Radio Astronomy Observatory (SARAO), which is a facility of the National Research Foundation, an agency of the Department of Science and Innovation. 
D.L. is a New Cornerstone Investigator. 
E. G. is supported by the Doctor Foundation of Qingdao Binhai University (No. BJZA2025025).

\item[Author Contributions] 
L.Z. and D.L. led the preparation of the manuscript, with contributions and feedback from all co-authors.
L.Z. discovered the source and served as Principal Investigator for the MeerKAT projects. A.R. acted as Technical Lead for the MeerKAT observations. A.R., F.C., M.B., and P.Z. supported the proposal, coordinated the execution of the observations, contributed to data processing, and assisted in refining the manuscript. E.G., W.C.G.H., and C.O.H. made significant contributions to the interpretation of the results. M.E.L. performed the Rotating Vector Model analysis and improved the geometric constraints on the source.
\end{addendum}

%######################## Methods ########################
\clearpage
\begin{methods}
\setcounter{figure}{0}
\setcounter{table}{0}
%\listoffigures
\captionsetup[figure]{labelfont={bf},labelformat={default},labelsep=period,name={Extended Data Fig.}}
\captionsetup[table]{labelfont={bf},labelformat={default},labelsep=period,name={Extended Data Tab.}}

\section*{MeerKAT Observations}  
Targeted radio pulsation searches are essential to determine whether CCOs are intrinsically radio-quiet or merely radio-faint, with direct implications for neutron-star magnetospheric physics and for evolutionary links between isolated neutron-star subclasses. We selected four well-established southern-hemisphere CCOs as targets for deep searches with MeerKAT. Their key parameters, including spin period (if known), period derivative, surface dipole magnetic field strength, associated SNR, SNR age, distance, and predicted DM, are listed in Extended Data Table~\ref{tab:CCOs_inf}. These sources were chosen for their precisely determined X-ray positions, absence of previous dedicated radio searches, and relatively small estimated distances ($\lesssim 2.5$\,kpc), all of which enhance sensitivity to faint pulsations.  

Each CCO was observed for 4\,h between December 2023 and January 2024 using the MeerKAT radio telescope in South Africa with the UHF receivers (544–1088\,MHz), centered at $f_{c}=816$\,MHz (Project ID: SCI-20230907-LZ-01). Data were recorded using the PTUSE (Pulsar Timing User Supplied Equipment) backend~\cite{Bailes2020} in pulsar-search mode, forming a single tied-array beam centered on the known source position, with a positional uncertainty corresponding to the $0.6'$ circle full-width at half-maximum (FWHM) of the MeerKAT UHF beam. The observation data were recorded with 4096 frequency channels across the 544\,MHz total bandwidth, coherently de-dispersed at the DM predicted by the YMW16 Galactic electron density model~\cite{Yao2017}, with a sampling interval of 120\,$\mu$s and 8 bits per sample, retaining all four Stokes parameters. A summary of the observations for the four CCO targets is given in Extended Data Table~\ref{tab:4CCOobs}.

Following the UHF detection of PSR~J1210$-$5226, we conducted an 8.5-hour follow-up observation on 18 October 2025 using MeerKAT director’s discretionary time (Project ID: DDT-20241205-LZ-01). This observation employed PTUSE in pulsar-search mode with the split-array configuration: 30 dishes at UHF and 30 dishes at L-band (856–1712\,MHz), enabling simultaneous dual-frequency coverage. An off-source beam was placed 5 arcminutes south of the target for Radio Frequency Interference (RFI) mitigation. Calibration was performed every 4.5 hours, and a short (2-minute) observation of the bright, polarized pulsar PSR~J1141$-$6545 was included to validate the instrumental setup. The full list of observations used in this work is listed in Extended Data Table~\ref{tab:1E1207obs}.

\section*{Periodicity and Single-Pulse Searches}  
As a first step, we used the \texttt{psrfits\_subband} routine from the \textsc{psrfits\_utils} package\footnote{\url{https://github.com/scottransom/psrfits_utils}} to sum the two polarizations, retaining total-intensity (Stokes~I) data only. During this process, the bandpass was flattened using the appropriate scale and offset values recorded during data acquisition.  

To improve sensitivity to possible pulse-nulling effects~\cite{Zhang2019a}, each 4\,h observation was subdivided into shorter segments of 2, 1, and 0.5\,h. The RFI was mitigated using standard \textsc{PRESTO} tools~\cite{Ransom2002} prior to dedispersion. For each target, dedispersion trials covered a DM range from 0 to 550\,pc\,cm$^{-3}$, encompassing the predicted values from the YMW16~\cite{Yao2017} and NE2001~\cite{Cordes2002} Galactic electron-density models, based on source position and estimated distance (see Extended Data Table~\ref{tab:4CCOobs} for the DM search ranges).  

For each de-dispersed time series, we carried out a blind Fourier-domain periodicity search using a \textsc{PRESTO}-based pipeline, optimized for fast-spinning pulsars including those in compact binaries. Since our targets have no evidence in favor of a binary nature, we performed \texttt{accelsearch} with $z_{\max}=0$. To improve sensitivity to narrow pulse profiles, where multiple harmonics contribute in the power spectrum, we summed up to 16 harmonics when evaluating candidate significance. 

Searching for long-period signals is particularly affected by red noise from system temperature variations, gain fluctuations, and sky background changes over long integrations~\cite{Lazarus2015, Singh2022}. To mitigate this, we employed the Fast Folding Algorithm (FFA), which is well-suited for detecting long-period pulsars~\cite{Morello2020, Zhou2024}. An FFA search was conducted using a \textsc{riptide}-based pipeline~\cite{Morello2020}. For the two targets without known spin periods, the search covered 0.1–100\,s with a running-median filter width of 5\,s for red-noise suppression. For the two targets with known periods, the search was restricted to a $\pm 0.2$\,s window around the known spin period, with a filter width of 1\,s. To avoid missing broad profiles, we allowed duty cycles between 0.001 and 0.5.  

From the 4-hour observation of 1E~1207.4$-$5209, we detected a 424-ms pulsar signal at a dispersion measure (DM) of 69\,pc,cm$^{-3}$. The detection was independently confirmed by both the Fourier-domain (\textsc{PRESTO}) and time-domain (\textsc{riptide}) periodicity search pipelines. Extended Data Figure~\ref{fig:FFA-FFT} presents the corresponding folded profiles from each pipeline. No significant periodic signal (${S/N} \geq 8$) was found in any search for the other three CCOs. 

We also performed a search for transient emission using a \textsc{heimdall}-based pipeline~\cite{Barsdell2012}, sensitive to single pulses such as those from Rotating Radio Transients (RRATs) or Fast Radio Bursts (FRBs). Trial DMs extended up to 5000\,pc\,cm$^{-3}$, and a boxcar matched-filter search was applied for pulse widths up to 60\,ms. No significant single-pulse events with ${\rm S/N} \geq 8$ were detected in any of the observations.

\section*{Sensitivity to pulsed radio emission}
The minimum detectable mean flux density in a targeted pulsar search is given by the modified radiometer equation~\cite{Lorimer2004}:
\begin{equation} 
S_{min}= \beta\frac{(\rm S/N)_{\rm min} T_{\rm sys}}{G\sqrt{n_{p}t_{\rm int}\Delta f}}\sqrt{\frac{W_{\rm eff}}{P-W_{\rm eff}}}, 
\label{eq:Smin}
\end{equation}
where $\beta$ is the digitization efficiency, we adopt $\beta=1.1$ for MeerKAT 8-bit sampling observation, S/N$_{\rm min}$ is the detection threshold, $T_{\rm sys}$ is the total system temperature, $G$ is the telescope gain, $n_p=2$ is the number of summed polarisations, $\Delta f$ is the effective (RFI-cleaned) bandwidth, $t_{\rm int}$ is the integration time, $P$ is the pulsar period, and $W_{\rm eff}$ is the effective pulse width. The effective width is computed as:
\begin{equation} 
W_{\rm eff} \;=\; \sqrt{W_i^2 \;+\; t_{\rm scatt}^2 \;+\; t_{\rm DM}^2 \;+\; t_{\rm samp}^2}\,
\end{equation}
where $W_i$ is the intrinsic width, $t_{\rm scatt}$ is the multipath scattering timescale, $t_{\rm DM}$ is the intra-channel dispersion smearing, and $t_{\rm samp}$ is the instrumental sampling time. A detailed discussion of these broadening terms and their impact on sensitivity is given by~\cite{Hessels2007}. For our observations with $\sim$60 MeerKAT antennas in the UHF band, we adopt $G = 2.6~{\rm K\,Jy^{-1}}$ and $T_{\rm sys} \simeq 31.5~{\rm K}$ (including receiver, sky, atmosphere, and spillover contributions; see~\cite{Ridolfi2022}). The usable bandwidth is $\Delta f = 500~{\rm MHz}$ after excising $\sim 8\%$ of channels affected by RFI, and the per-target integration time is $t_{\rm int}=4.0~{\rm h}$. 

The $S/N = 8$ radio flux upper limits for all four sources were estimated using Equation~\ref{eq:Smin} and are shown in Extended Data Figure~\ref{fig:Smin}. Although scattering may introduce additional pulse broadening, the sensitivity is found to depend primarily on the DM and the intrinsic duty cycle ($W_{i}/P$). Intra-channel dispersive smearing was largely mitigated by the use of coherent dedispersion in our observations. Adopting an $S/N = 8$ detection threshold, the radiometer equation yields flux upper limits of $\sim9~\mu{\rm Jy}$ for the four CCOs at the central frequency $f_{c} = 816$\,MHz, assuming an intrinsic duty cycle of 10\%, spin periods $\gtrsim 0.1$\,s, and DMs as listed in Extended Data Table~\ref{tab:4CCOobs}.

A comparison between search techniques highlights the need for a more realistic radiometer equation that includes the search efficiency factor $\varepsilon$. For the practical efficiencies of the FFA search ($\varepsilon_{\rm FFA} = 0.93$) and the FFT search with incoherent harmonic summing ($\varepsilon_{\rm FFT} = 0.70$ for a median pulsar duty cycle)~\cite{Morello2020}, the minimum detectable flux densities for our MeerKAT UHF observations are $\sim 10~\mu{\rm Jy}$ for the FFA search and $\sim 13~\mu{\rm Jy}$ for the FFT search.

We used the parameters described above in the following equation to calculate the upper limit of the flux density of a single pulse~\cite{Cordes2003}:
\begin{equation}
S_{\rm pulse}=\beta \frac{(\rm S/N)_{\rm min}T_{\rm sys}}{G\sqrt{n_{\rm p}W_{\rm obs}\Delta f}},
\end{equation}
where $W_{\rm obs}$ is the observed width of the burst. The corresponding fluence limits for assumed pulse widths of 60, 30, and 1\,ms are listed in Extended Data Table~\ref{tab:4CCOobs}.

\section*{Radio emission properties of PSR~J1210$-$5226}
We analyzed the 4-hour MeerKAT UHF observation to characterize the radio emission properties of PSR~J1210$-$5226. The data were folded using the best available ephemeris with the \textsc{dspsr} package\footnote{\url{http://dspsr.sourceforge.net}}~\cite{vanS11}
. The RFI was excised in both time and frequency using the \texttt{paz} routine from the \textsc{psrchive} suite\footnote{\url{http://psrchive.sourceforge.net}}~\cite{Hotan04}. Parallactic-angle corrections were applied with \texttt{pac}, and full-Stokes polarization calibration followed standard procedures~\cite{vanS10}, including the astronomical convention for defining the Stokes parameters. The fractional linear polarization and the position angle (PA) swing of the linear component were calculated following the procedure described by~\cite{Zhang2025}. 

From this dataset, we obtained a signal-to-noise ratio (S/N) of $\sim$30 after RFI excision, and measured a rotation measure (RM) of $+10 \pm 1$ rad m$^{-2}$ using \texttt{rmfit}. The pulse widths at 50\% and 10\% of the peak intensity, $W_{50}$ and $W_{10}$, were obtained from noise-free templates generated with \texttt{paas} and inspected using \texttt{pdv}. Using the radiometer equation and the known system parameters, we estimate a mean flux density of  $S_{\rm 816MHz} \simeq 33$\,$\mu$Jy for the 5 January 2024 detection in the UHF band. A summary of the derived radio parameters from this work, alongside previously reported X-ray measurements~\cite{Perez2025}, is presented in Table~\ref{tb:1E1207_par}.

The follow-up observation on 2025 October 18 was conducted in simultaneous dual-frequency (split-array) mode, with $\sim$30 MeerKAT antennas allocated to each of the UHF (544–1088\,MHz) and L-band (856–1712\,MHz) receivers. For both bands, we assumed a system equivalent gain of $G = 1.3~{\rm K\,Jy^{-1}}$, 8.5 hours of integration time, and a pulse duty cycle of 9\%. For L-band, we used an effective bandwidth of 650\,MHz and a system temperature of 26\,K~\cite{Ridolfi2022}. For UHF, system parameters were as described earlier. The total intensity profiles yielded S/N of $\sim$9 in the UHF band and $\sim$13 in L-band. Applying the radiometer equation, we estimated mean flux densities of $S_{\rm 816MHz} \simeq 13$\,$\mu$Jy and $S_{\rm 1284MHz} \simeq 14$\,$\mu$Jy in the UHF and L-band, respectively. Apparent flux variability across the observation is consistent with modulation by interstellar scintillation~\cite{Gitika2023} (see Extended Data Figure~\ref{fig:DDT-FFT}).

The polarization position angle (PA) curve of the linear polarization is shown in the top panel of the bottom plot of Figure~\ref{fig:1E1207_PolProf}. For an assumed dipole magnetic field geometry, the PA sweep traces the projected magnetic-field direction of the pulsar on the plane of the sky. The sweep can be fit as a function of rotation phase ($\phi$) using the rotating vector model (RVM)~\cite{Radhakrishnan1969} as 
\begin{equation}
    \tan({\rm PA} - \rm {PA}_{0}) = \frac{\sin\alpha \sin(\phi - \phi_{0})}{\sin(\alpha + \beta) \cos\alpha - \cos(\alpha + \beta) \sin\alpha \cos(\phi - \phi_{0})},
\end{equation}
where $\alpha$ is the angle between the rotational and magnetic axes of the pulsar, $\beta$ is the angular offset of the magnetic axis from our line of sight, and ${\rm PA}_{0}$ is the inflection point of the PA sweep which occurs at a reference rotation phase of $\phi_{0}$.
We sampled the RVM parameters using the Bayesian parameter estimation code \texttt{bilby} \cite{Ashton2019} as a wrapper for the \texttt{dynesty} nested sampling algorithm \cite{Speagle2020}.
In Figure~\ref{fig:RVMfit} we show the recovered two-dimensional posterior distribution for $\alpha$ and $\beta$.
The relatively low S/N of the profile and limited pulse phase range covered by the PA swing results in relatively unconstrained values of $\alpha = {82.3^{\circ}}^{+41.7}_{-38.8}$ and $\beta = {-18.6^{\circ}}^{+7.3}_{-7.2}$, corresponding to a viewing angle from the rotation axis of $\zeta = {59.3^{\circ}}^{+46.2}_{-33.0}$. 
Assuming a filled emission beam, we use the relation between the beam opening angle and the emission altitude to derive lines of constant emission height, shown in Figure~\ref{fig:RVMfit}, as
\begin{equation}
    \cos\rho = \cos\alpha \cos(\alpha + \beta) + \sin\alpha \sin(\alpha + \beta) \cos(W/2),
\end{equation}
where we assumed $W = W_{50}$, and
\begin{equation}
    \rho = 3\sqrt{\frac{\pi h_{\rm em}}{2 Pc}},
\end{equation}
in which $c$ is the speed of light in vacuum.
For standard pulsar emission heights of $\lesssim 1000$\,km \cite{Johnston2023}, the magnetic and viewing geometry are consistent with values of $\alpha \lesssim 53^{\circ}$ and $\beta \lesssim -17^{\circ}$ at the 1-$\sigma$ confidence interval, restricting the magnetic and rotational axes of PSR~J1210$-$5226 to a region of parameter space that is occupied by both aligned and oblique rotators.

\section*{Possible Physical Mechanisms for Radio Emission from 1E~1207.4$-$5209}
Our radio detection of 1E~1207.4$-$5209 offers new insights into the magneto-thermal evolution of CCOs. The inferred surface magnetic field and radio detectability are difficult to reconcile with a purely ``anti-magnetar” interpretation, in which the external dipole field is intrinsically weak and static. Although the dipole field inferred from spin-down is relatively low, these objects are not necessarily located below the conventional radio death line, suggesting that additional factors, such as magnetic-field geometry, magnetospheric configuration, and beaming, likely influence their radio emission. Moreover, the observed X-ray luminosity and surface temperature suggest the presence of additional magnetic energy, likely hidden below the crust.

One compelling possibility is that 1E~1207.4$-$5209 harbors a tangled magnetic field generated by a stochastic dynamo during the proto-neutron star phase~\cite{Gourgouliatos2020, Igoshev2021}. In this framework, the large-scale dipole field may be weak or absent, while toroidal components are confined to the crust and emerge on the surface as strong multi-polar magnetic structures with coupled magneto-thermal evolution. Such configurations can produce localized surface hot spots through Hall drift-assisted Ohmic decay of the magnetic field, consistent with the thermal X-ray emission and unusual absorption features observed in CCOs. Depending on the topology and strength of the magnetic field, certain magnetic structures may also favor coherent radio emission. Thus, the presence or absence of detectable radio pulsations among CCOs may be determined primarily by magnetic-field geometry and beaming configuration, rather than by field strength alone.

Another viable scenario is magnetic-field burial and subsequent re-emergence following fallback accretion in the immediate aftermath of the supernova explosion~\cite{Ho2011, Vigano2012, Torres-Forne2016}. In this model, the dipole field is temporarily submerged by a large amount of accreted material soon after the supernova~\cite{Chevalier1989,Bernal2010} and emerges over $10^3$–$10^5$ years \cite{Ho2011,igoshev2016} as the field diffuses out. 1E~1207.4$-$5209 may have reached the critical evolutionary stage where a previously buried dipole field has re-strengthened enough to support magnetospheric pair cascades and radio emission. Its relatively long spin period, possibly older evolutionary age, and—in contrast to the other two X-ray pulsating CCOs-its significantly larger inferred dipole magnetic field ($9.8\times10^{10}$ G versus 2.9 and $3.1\times10^{10}$ G)~\cite{Perez2025} all lend support to this hypothesis.

However, observational challenges remain. Long-term X-ray timing reveals that the spin-down rate decreased—not increased—after a glitch in 2015: from $(-1.232 \pm 0.007) \times 10^{-16}\,\mathrm{Hz\,s^{-1}}$ between 2002–2014 to $(-1.12 \pm 0.03) \times 10^{-16}\,\mathrm{Hz\,s^{-1}}$ between 2016–2023~\cite{Gotthelf2018, Perez2025}. Such behavior argues against a simple picture of field growth after burial. In addition, the persistent energies of the absorption lines suggest that the surface magnetic field has remained largely unchanged~\cite{Gotthelf2018}. It is possible, however, that the 2015 glitch triggered a reconfiguration in the magnetosphere or crust that enabled radio emission without altering the surface field strength. Glitches can either change the orientation of the magnetic dipole axis with respect to the spin axis \cite{ruderman1998,zhou2023} or temporally replenish some formerly passive magnetospheric region close to the surface with copious charged-pairs through the release of seismic and dissipated mechanical energies \cite{akbal2015,bransgrove2020}. The long-period sinusoidal-like variations seen in the timing residuals after the glitches in 1E~1207.4$-$5209 \cite{Gotthelf2020} are consistent with vortex oscillations driven by crustal displacements during a glitch \cite{erbil2023}. Since the changes caused by the glitch would have shorter duration of relaxation compared to the magnetic field evolution, if the radio emission is caused by the glitch it is expected to be of a transient nature.

Taken together, these models suggest that 1E~1207.4$-$5209 may represent a rare transitional phase within the CCO population—either due to evolutionary stage related to internal magnetic configuration, spin glitches, or a combination of both. Whether it continues to emit in radio or fades again will place important constraints on the physical mechanisms at play. Long-term radio monitoring, coupled with deeper, systematic searches of other CCOs, will be critical to disentangle the roles of magnetic-field geometry and time-dependent field evolution.

\clearpage
\begin{table*}
\centering
\caption{List of CCOs observed in this work, along with their key parameters: X-ray spin period ($P$), spin period derivative ($\dot{P}$), surface dipole magnetic field strength ($B_{s}$), name and age of the host SNR, distance to the SNR (Dis) from the CCO catalog$^{\ref{web:CCOs}}$ and estimated dispersion measure (DM${\mathrm{est}}$) derived from Galactic electron density models (YMW16\cite{Yao2017}, NE2001\cite{Cordes2002}).}
\setlength{\tabcolsep}{0.6mm}{
\label{tab:CCOs_inf}
\begin{tabular}{lccccccc}\hline
CCO                       & $P$    & $\dot{P}$     & $B_{s}$      & SNR             & SNR Age  & Dis   & DM$_{\rm{est}}$\\
                          & (s)    & ($10^{-18}$)  & ($10^{10}$G) &                 & (kyr)    & (kpc) & (pc cm$^{-3}$) \\\hline
RX~J0822.0$-$4300$^{a}$   & 0.112  & 9.28          & 2.9          & Puppis A        & 4.5      & 2.2   & 157, 417       \\
CXOU~J085201.4$-$461753   & ...    & ...           & ...          & G266.1-1.2      & 1        & 1     & 156, 214       \\
1E~1207.4$-$5209$^{a}$    & 0.424  & 22.2          & 9.8          & PKS 1209-51/52  & 7        & 2.2   & 127, 82        \\
1WGA~J1713.4$-$3949       & ...    & ...           & ...          & G347.3-0.5      & 1.6      & 1.3   & 52, 56         \\\hline
\multicolumn{8}{l}{$^{a}$ Two of three well-established X-ray pulsars in the class of CCOs.}\\ 
\end{tabular}}
\end{table*}

%RX J0822.0-4300
%1E 1207.4-5209
%CXOU J185238.6+004020

\begin{table*}
\footnotesize
\centering
\caption{Summary of MeerKAT observations and radio flux upper limits for the four CCO targets. The DM values for the four CCOs were estimated using the Galactic electron-density model YMW16~\cite{Yao2017} for coherent dedispersion and to calculate sensitivity limits. $S_{\rm min}$ denotes the upper limit on periodic pulsations at the central frequency $f_{c} = 816$\,MHz, assuming an intrinsic duty cycle of 10\% and spin periods $\gtrsim 0.1$\,s, without applying corrections for the practical efficiencies of the FFA search (0.93) and the FFT search with incoherent harmonic summing (0.70 for a median pulsar duty cycle)~\cite{Morello2020}. $S_{\rm sp}$ denotes the upper limit on single-pulse detections, with subscripts 60, 30, and 1 indicating assumed pulse widths of 60, 30, and 1\,ms, 
respectively.}
\label{tab:4CCOobs}
\renewcommand{\arraystretch}{1.0}
\setlength{\tabcolsep}{0.5mm}{
\begin{tabular}{lcccccccccc}\hline 
CCO & Observation & R.A.    & Decl.                   & Number of & \multicolumn{1}{c|}{DM$_{est}$}     & $S_{\rm min}$ & $S_{\rm sp,60}$ & $S_{\rm sp,30}$ & $S_{\rm sp,1}$\\
    & Date (UTC)  & (h:m:s) & ($^{\circ}$:$'$:$''$)   & antennas  & \multicolumn{1}{c|}{(pc cm$^{-3}$)} & ($\mu$Jy)     & (mJy ms) & (mJy ms) & (mJy ms)\\\hline 
RX~J0822.0$-$4300       & 2024-01-12  &  08:21:47.36 & $-$43:00:17.07 & 61 & \multicolumn{1}{c|}{157} & \multirow{4}{*}{9} & \multirow{4}{*}{14} & \multirow{4}{*}{19} & \multirow{4}{*}{105}\\
CXOU~J085201.4$-$461753 & 2024-01-05  &  08:52:01.40 & $-$46:17:53.0  & 61  & \multicolumn{1}{c|}{156} & & & & &\\
1E~1207.4$-$5209        & 2024-01-05  &  12:10:00.91 & $-$52:26:28.4  & 60  & \multicolumn{1}{c|}{127} & & & & &\\
1WGA~J1713.4$-$3949     & 2023-12-24  &  17:14:25.54 & $-$39:49:34.3  & 61  & \multicolumn{1}{c|}{52}  & & & & &\\
\hline         
\end{tabular}}
\end{table*}

\begin{table*}
\caption{Summary of MeerKAT observations of PSR~J1210$-$5226 used in this study. All data were recorded in pulsar-search mode with the beam pointed at the known X-ray position of the CCO 1E~1207.4-5209 (RA = 12$^\mathrm{h}$10$^\mathrm{m}$00.91$^\mathrm{s}$, Dec = $-52^\circ$26$'$28.4$''$).}
%\scriptsize
\footnotesize
\label{tab:1E1207obs}
\renewcommand{\arraystretch}{1.0}
\setlength{\tabcolsep}{0.5mm}{
\begin{tabular}{lcccccccccc}
\hline
Observation & Beam & Number of & Number of & Duration & DM             & Sampling time & Central Frequency & Bandwidth & Project\\
Date (UTC)  &      & antennas  & channels  & (hours)  & (pc cm$^{-3}$) & ($\mu$s)      & (MHz)             & (MHz)     & ID\\\hline 
2024-01-05  & 1    & 60        & 4096      & 4.0      & 127.0          & 120.47        & 815.93            & 544       & SCI-20230907-LZ-01\\
2025-10-18  & 2    & 31        & 256       & 8.2      & 69.0           & 240.94        & 815.93            & 544       & DDT-20251007-LZ-01\\
            &      & 30        & 256       & 8.2      & 69.0           & 306.24        & 1283.89           & 856       & DDT-20251007-LZ-01\\
\hline
\end{tabular}}
\end{table*}

\begin{figure*}
\begin{center}
\includegraphics[height=11cm,width=17cm]{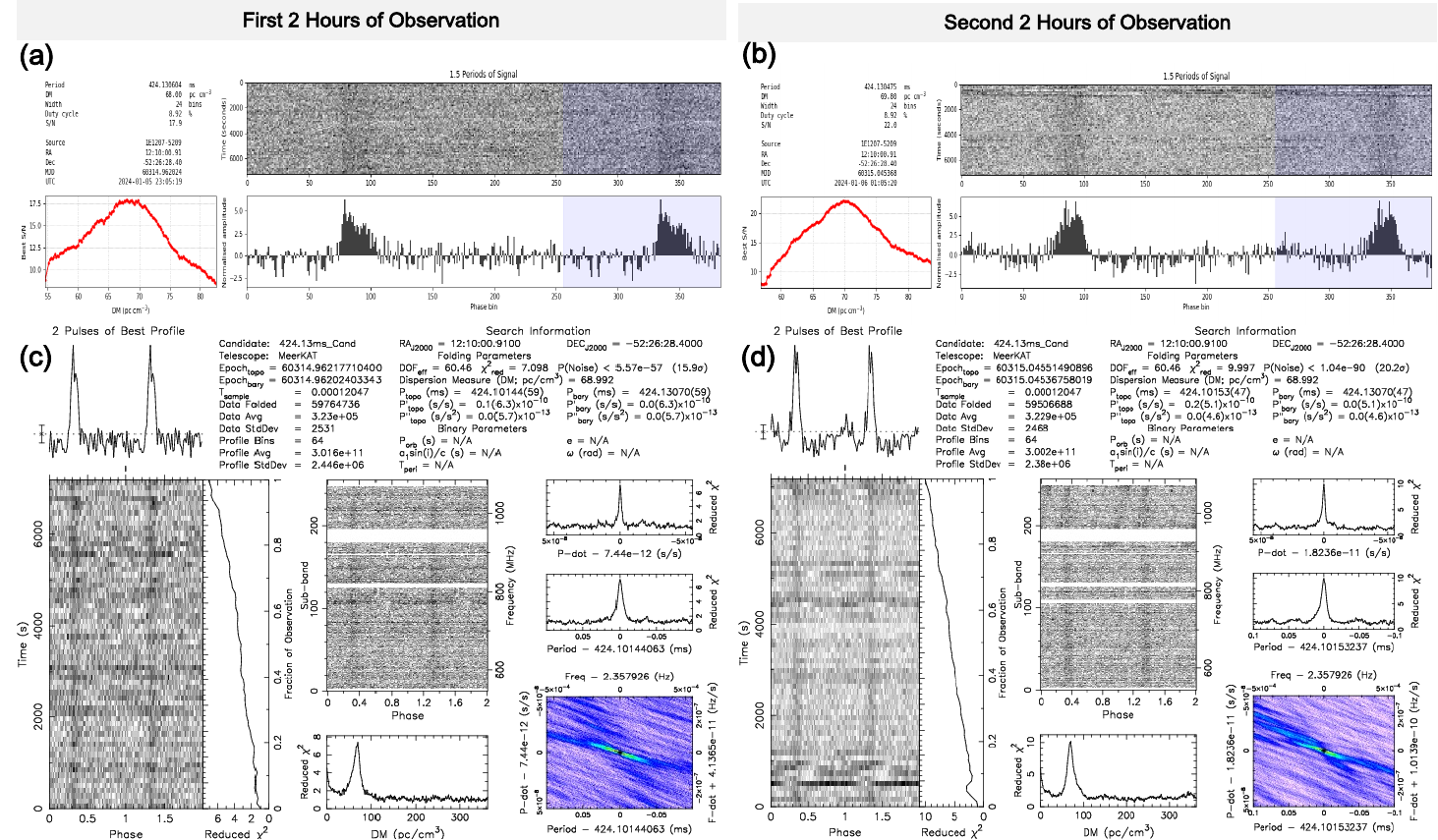}
\end{center}
\caption{
Discovery diagnostic plots of PSR~J1210$-$5226 from a 4-hour MeerKAT observation in the UHF band (544–1088\,MHz), centered at 816\,MHz and conducted on 5 January 2024. {\bf Upper panel:} (a) and (b) show results from the time-domain periodicity search using the \textsc{riptide}-based pipeline, applied to two consecutive 2-hour segments of the observation. {\bf Lower panel:} (c) and (d) show results from the Fourier-domain periodicity search using the \textsc{PRESTO}-based pipeline, applied to the same two segments.
}
\label{fig:FFA-FFT}
\end{figure*}

\begin{figure*}
\begin{center}
\includegraphics[width=0.98\textwidth]{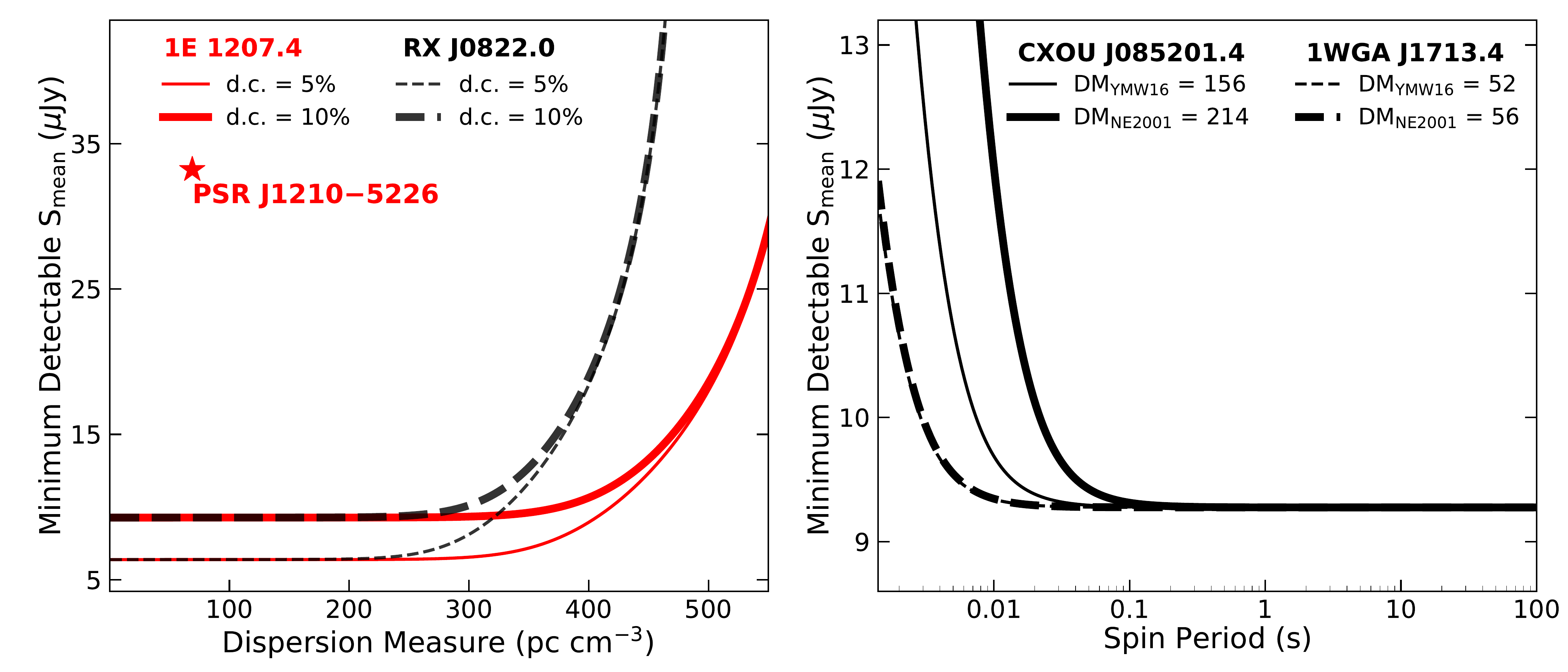}
\end{center}
\caption{{\bf Sensitivity curves for the four CCOs observed with a 4-hour MeerKAT integration in the UHF band (544–1088\,MHz).}
{\bf Left:} Theoretical upper limits on radio flux density for the two CCOs with known X-ray spin periods, plotted as a function of dispersion measure (DM) for various assumed intrinsic duty cycles (d.c.). The red star marks PSR~J1210$-$5226 at its observed DM and measured mean flux density from this work. The predicted DMs from Galactic electron density models are 127 and 82\,pc cm$^{-3}$ (YMW16, NE2001, respectively) for 1E~1207.4$-$5209, and 157 and 417\,pc cm$^{-3}$ for RX~J0822.0$-$4300.
{\bf Right:} Theoretical flux density limits for the two CCOs without known spin periods, shown as a function of spin period, assuming a 10\% intrinsic duty cycle. In all cases, dispersive smearing is the dominant source of pulse broadening and sensitivity degradation.}
\label{fig:Smin}
\end{figure*}

\begin{figure*}
\begin{center}
\includegraphics[height=14cm,width=16cm]{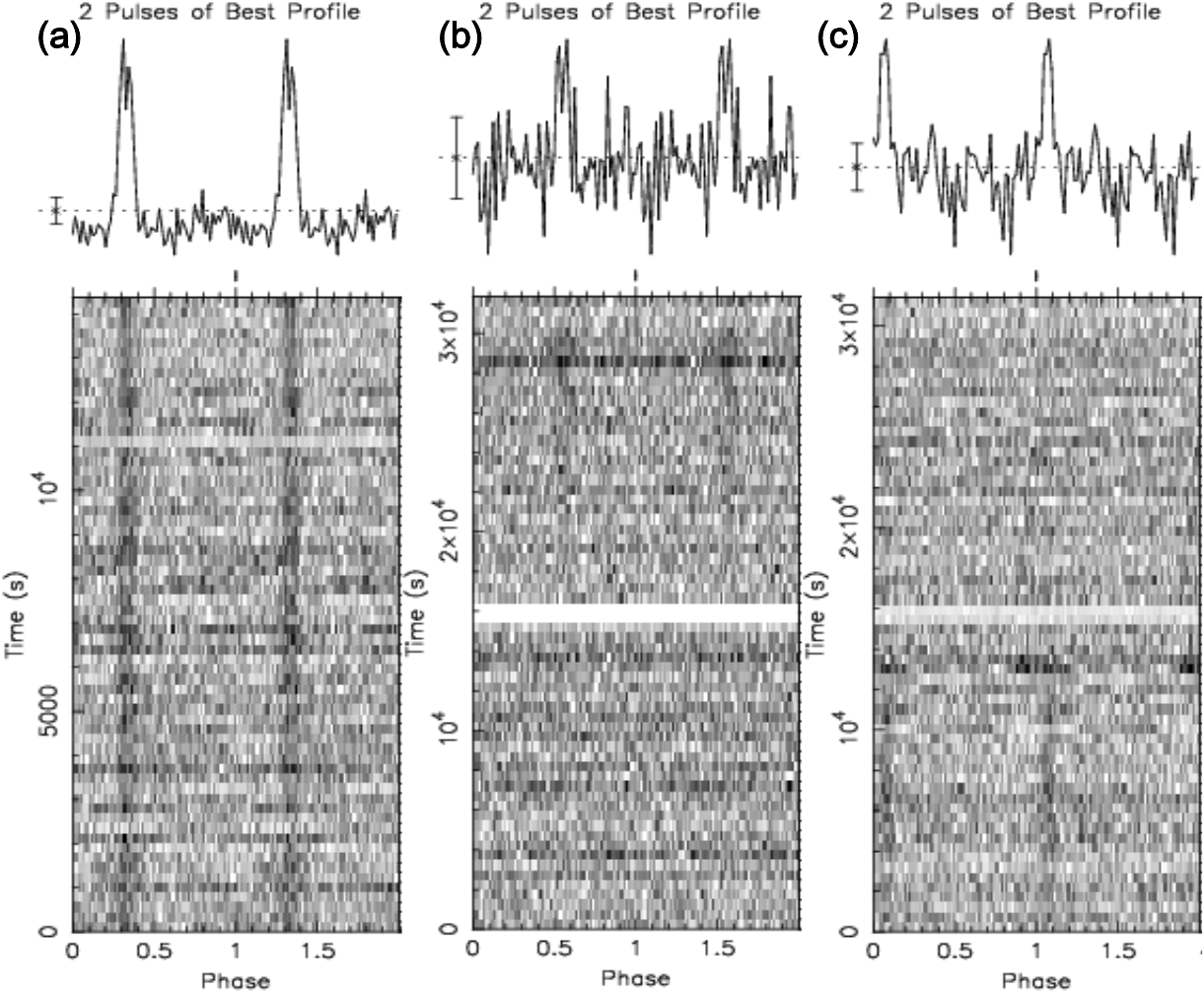}
\end{center}
\caption{
Intensity as a function of time and spin phase for the two MeerKAT observations of PSR~J1210$-$5226.
{\bf(a)} Observation on 2024-01-05 using 60 dishes in the UHF band (544–1088\,MHz).
{\bf(b)} Observation on 2025-10-18 using 31 dishes in the UHF band.
{\bf(c)} Observation on 2025-10-18 using 30 dishes in the L-band (856–1712\,MHz).
White horizontal stripes mark intervals excised due to radio frequency interference. The flux-density variations across time are consistent with interstellar scintillation~\cite{Gitika2023}.}
\label{fig:DDT-FFT}
\end{figure*}

\begin{figure*}
\begin{center}
\includegraphics[height=15cm,width=14cm]{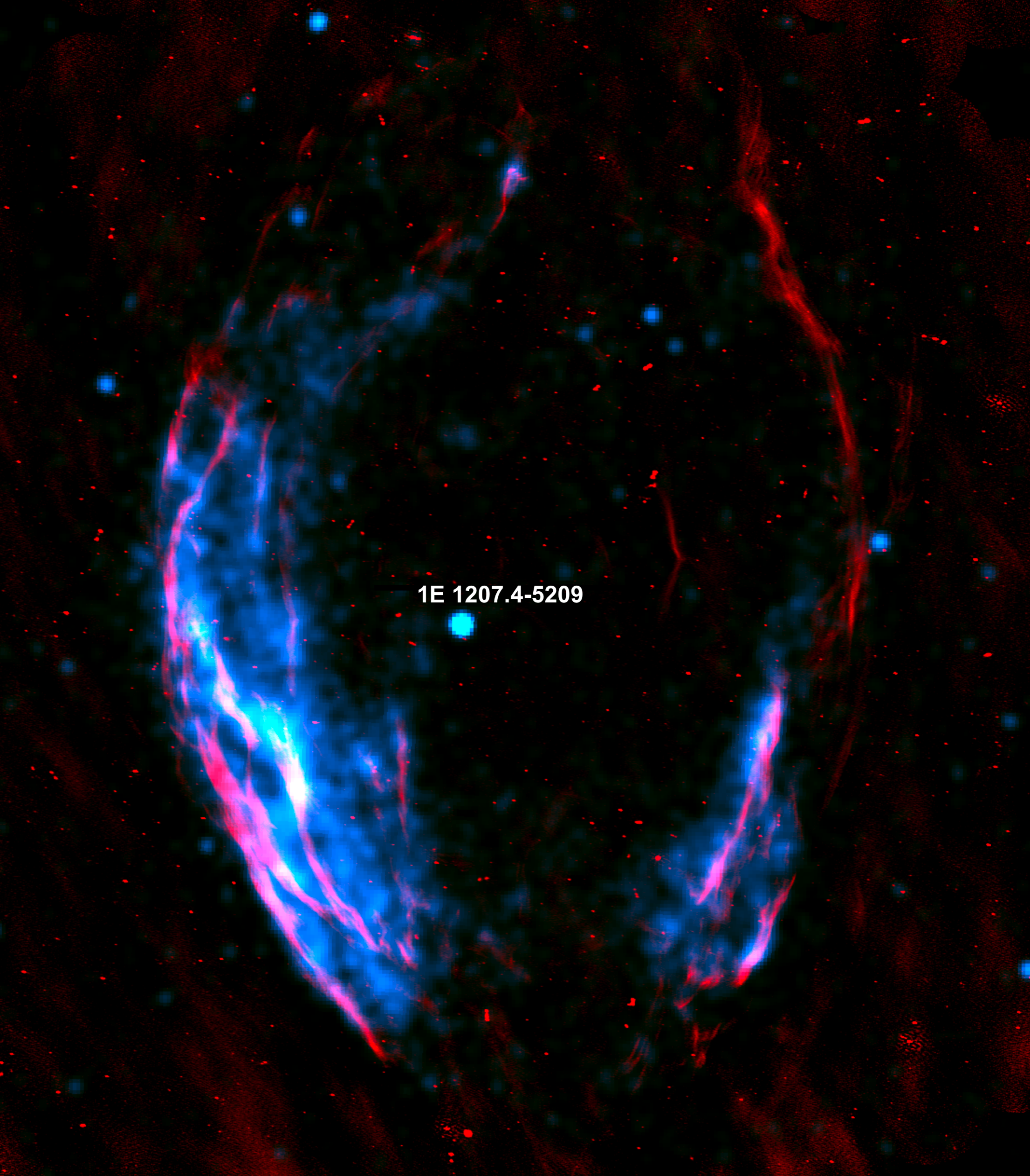}
\end{center}
\caption{Composite image of the supernova remnant PKS~1209$-$51/52, combining radio data from MeerKAT (1.3\,GHz \cite{Cotton2024}; shown in red) and X-ray data from eROSITA (0.4--2.0\,keV; shown in cyan). The central compact object 1E~1207.4$-$5209 is prominently visible at the centre of the remnant.}
\label{fig:MKT-Xray}
\end{figure*}

\clearpage
\subsection{Data availability}
Our processed data collection is publicly available via the Science Data Bank\footnote{\url{https://doi.org/10.57760/sciencedb.31405}}.
The MeerKAT raw data used in this study are available via the SARAO archive\footnote{\url{https://archive.sarao.ac.za}} under project ID SCI-20230907-LZ-01 and DDT-20251007-LZ-01.

\subsection{Code availability}
Specific scripts used in the data analysis are available on request from L. Z. The standard data reduction packages are available at their respective websites: \\
\textsc{presto} - \url{https://github.com/scottransom/presto};\\
\textsc{riptide} - \url{https://github.com/v-morello/riptide};\\
\textsc{DSPSR} - \url{http://dspsr.sourceforge.net};\\
\textsc{PSRCHIVE} - \url{http://psrchive.sourceforge.net};\\
\textsc{TEMPO} - \url{http://tempo.sourceforge.net};\\
\textsc{TEMPO2} - \url{https://sourceforge.net/projects/tempo2}.\\
\end{methods}

%\bibliographystylemd{naturemag}
%\bibliographymd{reference-md}

\end{document}